\documentclass[onecolumn]{aastex62}

\submitjournal{ApJ}

\shorttitle{Current sheets in near-Sun solar wind}
\shortauthors{A. Lotekar}

\usepackage{multirow}

\usepackage{color}
\newcommand{\blue}{\textcolor{black}}
\usepackage{lineno}

\begin{document}

\title{Kinetic-scale current sheets in near-Sun solar wind: properties, scale-dependent features and reconnection onset}

\correspondingauthor{A. Lotekar}
\email{ablotekar@gmail.com}

\author{A. Lotekar}
\affil{Swedish Institute of Space Physics, Uppsala, Sweden}

\author{I.Y. Vasko}
\affil{Space Sciences Laboratory, University of California at Berkeley}
\affil{Space Research Institute of Russian Academy of Sciences, Moscow, Russia}

\author{T. Phan}
\affil{Space Sciences Laboratory, University of California at Berkeley}

\author{S.D. Bale}
\affil{Space Sciences Laboratory, University of California at Berkeley}
\affil{Department of Physics, University of California at Berkeley}

\author{T.A. Bowen}
\affil{Space Sciences Laboratory, University of California at Berkeley}

\author{J. Halekas}
\affil{The University of Iowa, Iowa, USA}

\author{A.V. Artemyev}
\affil{University of California, Los Angeles, USA}
\affil{Space Research Institute of Russian Academy of Sciences, Moscow, Russia}

\author{Yu. Khotyaintsev}
\affil{Swedish Institute of Space Physics, Uppsala, Sweden}

\author{F.S. Mozer}
\affil{Space Sciences Laboratory, University of California at Berkeley}
\affil{Department of Physics, University of California at Berkeley}






\begin{abstract}
We present statistical analysis of 11,200 proton kinetic-scale current sheets (CS) observed by Parker Solar Probe during 10 days around the first perihelion. The CS thickness $\lambda$ is in the range from a few to 200 km with the typical value around 30 km, while current densities are in the range from 0.1 to 10 $\mu {\rm A/m^2}$ with the typical value around 0.7 $\mu {\rm A/m^2}$. These CSs are resolved thanks to magnetic field measurements at 73--290 Samples/s resolution. In terms of proton inertial length $\lambda_{p}$, the CS thickness $\lambda$ is in the range from about $0.1$ to $10\lambda_{p}$ with the typical value around 2$\lambda_{p}$. The magnetic field magnitude does not substantially vary across the CSs and, accordingly, the current density is dominated by the magnetic field-aligned component. The CSs are typically asymmetric with statistically different magnetic field magnitudes at the CS boundaries. The current density is larger for smaller-scale CSs, $J_0\approx 0.15 \cdot (\lambda/100\;{\rm km})^{-0.76}$ $\mu {\rm A/m^2}$, but does not statistically exceed the Alfv\'en current density $J_A$ corresponding to the ion-electron drift of local Alfv\'{e}n speed. The CSs exhibit remarkable scale-dependent current density and magnetic shear angles, $J_0/J_{A}\approx 0.17\cdot (\lambda/\lambda_{p})^{-0.67}$ and $\Delta \theta\approx 21^{\circ}\cdot (\lambda/\lambda_{p})^{0.32}$. Based on these observations and comparison to recent studies at 1 AU, we conclude that proton kinetic-scale CSs in the near-Sun solar wind are produced by turbulence cascade and they are automatically in the parameter range, where reconnection is not suppressed by the diamagnetic mechanism, due to their geometry dictated by turbulence cascade.
\end{abstract}

\keywords{solar wind --- plasmas --- waves---current sheet}



\section{Introduction}

Spacecraft measurements in the solar wind allow {\it in-situ} analysis of turbulence in a magnetized weakly-collisional plasma that is typical of numerous astrophysical systems \citep[e.g.,][]{MacLow99,Matthaeus&Velli11,Zhuravleva14,Li21:apjl}. \blue{Remote and {\it in-situ} measurements} at radial distances larger than 0.3 au showed that solar wind heating should continuously occur within a few tens of solar radii of the Sun as well as further out in the heliosphere \citep[e.g.,][]{Kohl96:apj,Cranmer09,Hellinger13:jgr}. \blue{The dissipation of turbulent magnetic field fluctuations is expected to be the most important solar wind heating mechanism} \citep[e.g.,][]{Vasquez07:turbulence_transfer,Cranmer09,Hellinger13:jgr}. Numerical simulations \blue{showed} that turbulence dissipation should be spatially intermittent with substantial plasma heating localized around coherent structures, such as current sheets, which occupy a relatively small volume \citep[][]{Karimabadi13,Zhdankin13:apjl,Zhdanin14:apj,Wan14:apj,Wan16:phpl}. Spacecraft measurements at 1 au confirmed that plasma heating indeed occurs around current sheets \citep[][]{Osman11:apjl,Osman12b,Wu13:apj}, but contributions of various plasma heating mechanisms are still not entirely understood \citep[e.g.,][]{Kiyani15,Goldstein15}. \blue{One of the mechanisms} initiating particle heating and turbulence dissipation is \blue{magnetic reconnection, which can also substantially} affect development of the turbulence cascade at proton and sub-proton scales \citep[e.g.,][]{Matthaeus&Lamkin86,Servidio11:npg,Servidio15:jpp,Cerri&Califano17,Franci17,Franci18,Papini19:apj}. Because magnetic reconnection is sufficiently fast only in current sheets with thickness around proton kinetic scales \citep[e.g.,][]{Cassak06:apj}, the understanding of turbulence dissipation in astrophysical plasma can be advanced \blue{by analysis of proton kinetic-scale current sheets observed} in the solar wind. In this paper we present a statistical analysis of proton kinetic-scale current sheets (CS) observed by Parker Solar Probe spacecraft in the previously unexplored near-Sun solar wind, within a few tens of solar radii from the Sun \citep{Fox16:ssr}.


The presence of CSs, originally termed directional discontinuities, on a wide range of temporal scales was established by previous spacecraft measurements at 1 au \citep[e.g.,][]{burlaga77,tsurutani79,lepping86,soding01,Artemyev18:apj,Artemyev19:jgr}. In most of these studies CSs were selected using magnetic field measurements with resolution of a few seconds at best. The typical CS thickness was around one thousand kilometers or ten proton inertial lengths, the occurrence rate was a few tens per day. However, magnetic field measurements at higher resolution (1/3 and 1/11 s) allowed resolving CSs with thickness around one proton inertial length and showed they are much more abundant\blue{, a few hundred CSs per day} \citep{Vasquez07,Podesta17:jgr,Vasko21a:apjl,Vasko21b:apjl}. The magnetic field variation across the proton kinetic-scale CSs is predominantly a rotation through some shear angle, rather than magnitude variation \citep{Vasquez07,Vasko21a:apjl,Vasko21b:apjl}, that is also typical of larger-scale CSs \citep[e.g.,][]{Burlaga1969:solphys,lepping86,Artemyev19:jgr}. \blue{Based on the distribution of waiting times, it was} hypothesized that kinetic-scale CSs in the solar wind are produced by turbulence cascade \citep{Vasquez07,Greco08,Greco09:apjl,Perri12:prl}. \cite{Vasko21b:apjl} has further supported this hypothesis by demonstrating that \blue{the current density and} shear angle across the CSs depend on CS thickness in a scale-invariant fashion expected for turbulent fluctuations.

The observations of reconnecting CSs at 1 au were reported fairly recently \citep[][]{Phan06:nature,Gosling07:grl_prevalence,Gosling07:apj}. The reconnection was identified by a plasma jet within CS and found to be often, though not always, associated with a bifurcated magnetic field profile \citep{Gosling&Szabo08,Phan10,Mistry15:grl}. The spacecraft measurements also showed that magnetic reconnection in the solar wind does result in plasma heating \citep{Phan06:nature,Enzl14:apj,Pulupa:2014,Mistry17:jgr}. The plasma measurements at 3s resolution or, equivalently, $\approx 1000$ km spatial resolution showed that magnetic reconnection at 1 au is relatively rare, about one reconnecting CS per day \citep{Phan10,Gosling12,Osman14:prl}. The fundamental \blue{questions are what is} the occurrence of reconnection at proton kinetic scales and why it is so rare at spatial scales of $\approx 1000$ km. Although the occurrence of reconnection at proton kinetic scales has not been established yet, the recent analysis by \cite{Vasko21a:apjl} has shown that \blue{the presence or absence of reconnection in such current sheets is not} determined by the diamagnetic suppression condition. Note that previously \cite{Phan10} showed that reconnecting CSs are in the parameter range, where reconnection cannot be suppressed by the diamagnetic mechanism \citep{Swisdak03:jgr,Swisdak_2010}, while the analysis by \citet{Vasko21a:apjl} showed that all proton kinetic-scale CSs are automatically in that parameter range due to their geometry dictated by turbulence cascade.

The high-resolution magnetic field measurements aboard the recently launched Parker Solar Probe spacecraft allow resolving proton and sub-proton kinetic-scale CSs in the near-Sun solar wind. The previous measurements of near-Sun solar wind, at radial distances as close as 60 solar radii, were done aboard Helios spacecraft \citep{Rosenbauer77}. However, the highest magnetometer resolution aboard Helios was 0.25s, and only CSs with thickness larger than about ten proton inertial lengths were resolved \citep{soding01}. \cite{Phan20:apjs} have recently used Parker Solar Probe measurements to address the occurrence of magnetic reconnection in the near-Sun solar wind. Using plasma measurements at 0.2--0.9s cadence and magnetic field measurements downsampled to 0.2s, they resolved CSs with thickness larger than about ten proton inertial lengths and found negligible amount of reconnecting CSs among them.


In this paper we present analysis of 11,200 proton kinetic-scale CSs observed by Parker Solar Probe around the first perihelion. \blue{The properties of these current sheets will be compared to those recently reported at 1 au \citep{Vasko21a:apjl,Vasko21b:apjl}.} The paper is organized as follows. Section \ref{sec1} describes the dataset and methodology. Section \ref{sec2} overviews the considered interval and presents several case studies. Section \ref{sec3} presents the epoch analysis and CS properties. Section \ref{sec4} presents the test of the diamagnetic suppression condition of magnetic reconnection. Section \ref{sec5} presents scale-dependence of various CS properties. Sections \ref{sec6} and \ref{sec7} present discussion and summary of the results.



\section{Data and methodology \label{sec1}}

We consider Parker Solar Probe (PSP) measurements over 10 days around the first perihelion, from November 1 to November 10, 2018. We use the magnetic field measurements provided by the FIELDS instrument suite \citep{Bale16:ssr}. Specifically, we use the merged fluxgate and search-coil magnetometer measurements \citep{Bowen20:jgr}, with sampling rate of about 290 S/s (Samples per second) from November 5 to November 7 and gradually reduced to 73 S/s as the spacecraft moved away from the perihelion. We use ion and electron moments provided the Solar Wind Electrons Alphas and Protons (SWEAP) instrument suite \citep{Kasper16:ssr}, including the Solar Probe Cup (SPC) \citep{Case_2020} and the
Solar Probe Analyzers-Electrons (SPAN-E) \citep{Whittlesey20}. The estimates of proton density, bulk velocity and radial proton temperature were all available at 0.2--0.9s cadence. The electron velocity distribution functions available at 28s cadence have been previously analyzed and fitted to a combination of core, halo and strahl populations \citep{Halekas20:apj}. We use the total electron density of the three components and core electron temperature as a proxy of the total electron temperature, because thermal and dynamic pressures of halo and strahl populations were less than a few percent of the core electron pressure \citep{Halekas20:apj}. We will also present plasma density estimates delivered by the quasi-thermal noise spectroscopy available at about 7s cadence \citep{Moncuquet20:apj}. The procedure of CS selection and methodology\blue{, briefly described below, are equivalent to those used at 1 au \citep{Vasko21a:apjl,Vasko21b:apjl}}.

The selection of CSs is based on the Partial Variance Increments (PVI) method \citep[e.g.,][]{Greco08,Greco18}. We compute PVI index, ${\rm PVI}(t,\tau)=\left(\sum_{\alpha}\Delta B_{\alpha}^2(t,\tau)/\sigma_{\alpha}^2\right)^{1/2}$, where $\Delta B_{\alpha}(t,\tau)=B_{\alpha}(t+\tau)-B_{\alpha}(t)$ are magnetic field increments of various magnetic field components ($\alpha=X,Y,Z$) and $\sigma_{\alpha}$ are standard deviations of $\Delta B_{\alpha}(t,\tau)$ computed \blue{over 1h intervals, that is over several outer correlation scales} of the turbulence in the near-Sun solar wind \citep{Chen20:apjs}. Coherent structures at various temporal scales correspond to non-Gaussian fluctuations with, for example, PVI$>$5. We used only PVI index computed at the minimum time increment $\tau$ determined by the sampling rate of magnetic field measurements, so that $\tau$ is in the range from 1/73 to 1/290 s. This allows analysis of the thinnest coherent structures still resolvable by PSP magnetic field measurements. Because PVI index is proportional to local current density \citep[][]{Chasapis17:J_PVI,Yordanova20:front}, our focus on fluctuations with ${\rm PVI}>5$ translates into selection of the most intense coherent structures. Note that it is not our purpose to select all CSs present in the solar wind at various spatial scales, but rather to collect a sufficiently representative dataset of proton kinetic-scale CSs to address their properties and origin. 
 
\begin{figure}[ht!]
\centering
\includegraphics[width=0.7\textwidth]{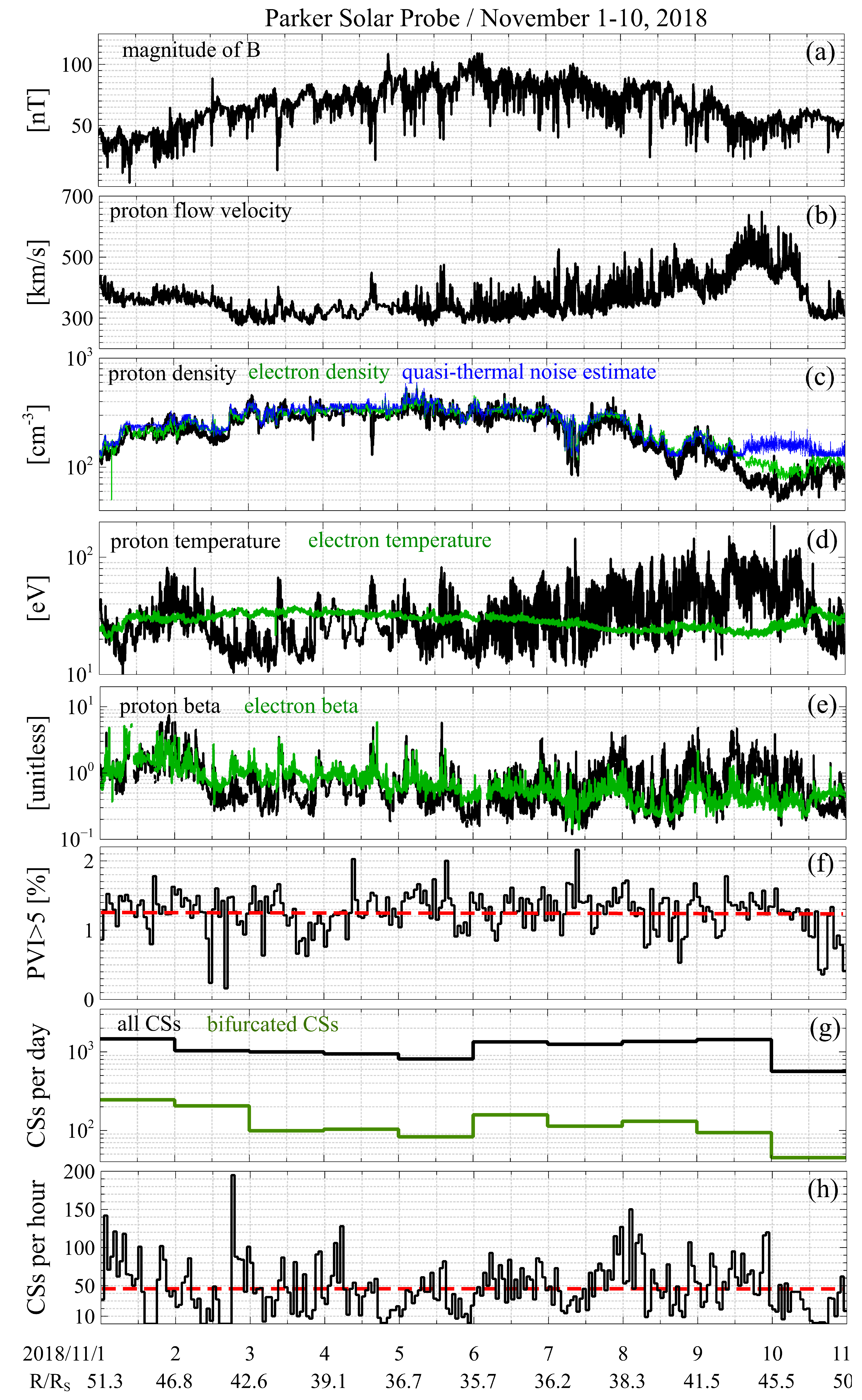}
\caption{Overview of Parker Solar Probe measurements over 10 days around the first perihelion, November 1 to 10, 2018. Panels (a)--(b) present 1 minute averages of the magnetic field magnitude, proton flow velocity, electron and proton densities along with plasma density estimates by the quasi-thermal noise spectroscopy, electron and proton temperatures and electron and ion betas. Panel (f) shows the percentage of fluctuations (relative number of magnetic field increments) with PVI index larger than 5. Panel (g) presents the total number of current sheets (CS) observed per day along with the number of CSs classified as bifurcated. Panel (h) shows the number of CSs observed per hour. The red lines in panels (f) and (h) represent averaged (over 10 days) values of the corresponding quantities.}
\label{fig1}
\end{figure}

In addition to current sheets, there are other types of coherent structures among the non-Gaussian fluctuations (PVI$>5$) such as Alfv\'{e}n vortexes \citep{Perrone20:apj} and ion-cyclotron waves \citep[][]{Bowen20,Bowen20:apj2}. We considered \blue{each continuous cluster of points} with PVI$>$5 over several nested 0.1--2s intervals around its center and \blue{used the Maximum Variance Analysis \citep[e.g.,][]{Sonnerup&Scheible98:issi} to compute unit vector $\textbf{\emph{x}}'$ along the magnetic field component with the largest variation}. We visually inspected all ${\bf B}\cdot \textbf{\emph{x}}'$ profiles and selected clusters of points with ${\bf B}\cdot \textbf{\emph{x}}'$ reversing the sign within at least one of the intervals. We then manually adjusted the boundaries, so that each boundary has at least ten points of magnetic field measurements, and excluded events with substantial relative variations of the magnetic field at the boundaries. The selected CSs were visually classified into non-bifurcated and bifurcated, the latter type often seen in reconnecting CSs \citep[e.g.,][]{Phan10,Phan20:apjs,Mistry17:jgr}. A short temporal duration of \blue{the CSs did} not allow establishing the presence or absence of reconnection jets using proton flow velocity measurements and, thus, we could not determine the fraction of reconnecting CSs in our dataset. The final dataset includes 11,200 CSs with 1,277 of them classified as bifurcated. 

For each CS we use local coordinate system $\textbf{\emph{xyz}}$ most suitable for describing a local CS structure \citep[e.g.,][]{Knetter04,Gosling&Phan13,Phan20:apjs}: unit vector $\textbf{\emph{z}}$ is along the CS normal determined by the cross-product of magnetic fields at the CS boundaries; unit vector $\textbf{\emph{x}}$ is along $\textbf{\emph{x}}'-\textbf{\emph{z}}\cdot (\textbf{\emph{x}}'\cdot \textbf{\emph{z}})$; unit vector $\textbf{\emph{y}}$ completes the right-handed coordinate system, $\textbf{\emph{y}}=\textbf{\emph{z}}\times \textbf{\emph{x}}$. Because the Taylor hypothesis was valid during the first perihelion \citep{Chen20:apjs,Chhiber21:apj}, the temporal profiles of the CSs are translated into spatial profiles by computing the spatial distance, $z=-V_{n}(t-t_0)$, where $t_0$ is an arbitrary moment of time and $V_{n}$ is the normal \blue{component of local proton flow} velocity at the moment closest to a CS. We estimate the current density components as follows
\begin{eqnarray}
 J_{x}=\frac{1}{\mu_0 V_{n}}\frac{dB_{y}}{dt},\;\;\; J_{y}=-\frac{1}{\mu_0 V_{n}}\frac{dB_{x}}{dt},
\label{eq:JxJy}
\end{eqnarray}
where $\mu_0$ is the vacuum permeability. We \blue{also present current densities} parallel and perpendicular to local magnetic field, $J_{||}=(J_{x}B_{x}+J_{y}B_{y})/B$ and  $J_{\perp}=(J_{y}B_{x}-J_{x}B_{y})/B$. By noting that magnetic field of a locally planar CS can be described as 
\begin{eqnarray}
{\bf B}=B(z)\sin \theta(z)\;\textbf{\emph{x}}+B(z)\cos\theta(z)\;\textbf{\emph{y}}+B_{z}\;\textbf{\emph{z}},
\label{eq:B}
\end{eqnarray}
we find that
\begin{eqnarray}
J_{||}=\frac{B}{\mu_0}\frac{d\theta}{dz},\;\;\; J_{\perp}=\frac{1}{\mu_0}\frac{dB}{dz},
\label{eq:JparJper}
\end{eqnarray}
where $\theta (z)$ and $B(z)$ describe respectively the magnetic field rotation and magnitude variation within the CS, and $B_{z}$ is negligibly small compared to $B(z)$. Note that Eq. (\ref{eq:B}) is the most general expression for the magnetic field of a CS with non-zero $B_y$, while specific models \blue{widely used in theoretical studies \citep[e.g.,][]{Landi15:apj,Boldyrev18,Neukirch20:apj} correspond to specific profiles of $B(z)$ and $\theta(z)$}. Eqs. (\ref{eq:JparJper}) show that the parallel current density determines magnetic field rotation, while perpendicular current density determines variation of the magnetic field magnitude within the CS. The ratio between perpendicular and parallel current densities can be approximately estimated as follows
\begin{eqnarray}
J_{\perp}/J_{||}\approx \Delta B/\langle B\rangle \Delta \theta
\label{eq:Jper_Jpar}
\end{eqnarray}
where $\langle B\rangle$ is typical/averaged magnetic field magnitude, $\Delta B$ and $\Delta \theta$ are respectively magnetic shear angle and variation of the magnetic field magnitude across CS.

The collected proton kinetic-scale CSs have relatively short temporal duration, more than 85\% of the CSs have temporal duration of less than 0.5s. The cadence of plasma measurements, especially 28s cadence for electrons, does not allow reliable estimates of plasma $\beta$ at the CS boundaries, because magnetic fields at the CS boundaries generally vary on a time scale of 0.9--28s. We determine the variation of plasma $\beta$ across each CS \blue{as proposed by \cite{Vasko21a:apjl}}. We assume there is a pressure balance, $8\pi P +B^2=8\pi \Pi={\rm const}$ or $8\pi P/B^2 +1=8\pi \Pi/ B^2$, where $P$ is the thermal plasma pressure. There are several proton measurements and one point of electron measurements around each CS, and the constant parameter $\Pi$ can be determined by averaging the pressure balance across the CS, $\Pi=\langle B^2\rangle/8 \pi+\langle P \rangle$ and $8\pi \Pi\langle 1/B^{2}\rangle=\langle 8 \pi P/B^2 \rangle+1$, where we assume that the spatial averaging of thermal plasma pressure is equivalent to time-averaging done by PSP plasma instruments. The variation of plasma beta can be \blue{estimated} as follows
\begin{eqnarray}
\Delta \beta\approx (1+\beta)\; \Delta \left(B^{-2}\right)\;/\;\langle B^{-2}\rangle\
\label{eq:dbeta}
\end{eqnarray}
where $\beta=\langle 8 \pi P/B^2 \rangle=(8\pi \langle P\rangle+\langle B^2\rangle) \langle B^{-2}\rangle-1$ is the averaged plasma beta. 



\begin{figure}[ht!]
\centering
\includegraphics[width=0.7\textwidth]{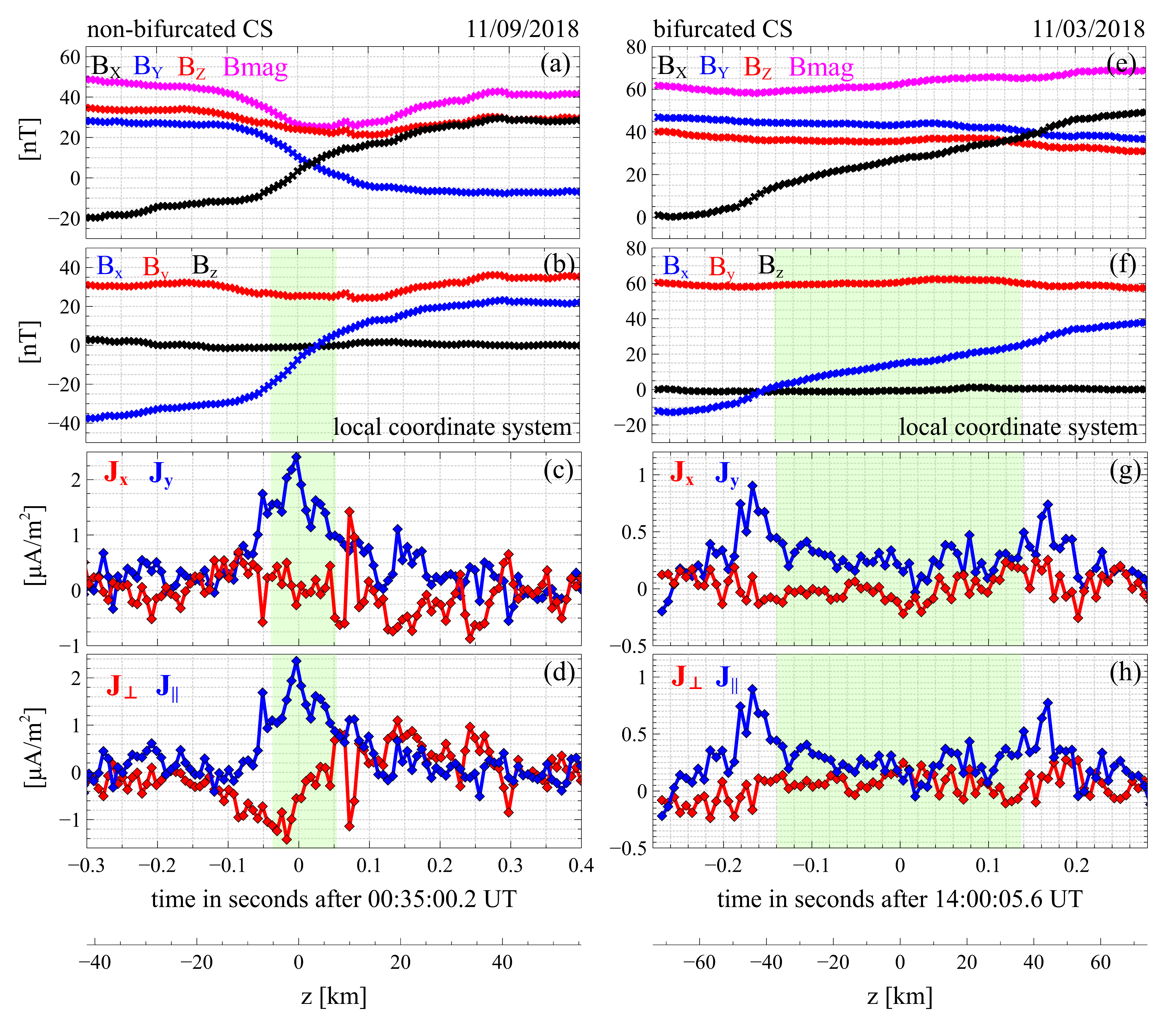}
\caption{The examples of non-bifurcated (left panels) and bifurcated (right panels) CSs from our dataset. Panels (a) and (e) present the magnetic field magnitude and three components in the spacecraft coordinates. Panels (b) and (f) show three magnetic field components in the local CS coordinate system defined in Section \ref{sec1}. Panels (c) and (g) present current density components $J_{x}$ and $J_{y}$ determined by Eqs. (\ref{eq:JxJy}), while panels (d) and (h) present current density components parallel and perpendicular to local magnetic field, $J_{||}=(J_{x}B_{x}+J_{y}B_{y})/B$ and $J_{\perp}=(J_{y}B_{x}-J_{x}B_{y})/B$. The spatial coordinate $z$ across each CS was computed as $z=-V_{n} (t-t_0)$ and shown at the bottom of panels (d) and (h), where $t_0$ is the time instant indicated in the temporal axes and $V_{n}$ is the normal component of the proton flow velocity measured at the moment closest to the CS. \blue{The region highlighted in panels (b)--(d) and (f)-(h) is the CS central region, where $|B_{x}-\langle B_{x}\rangle|<0.2\Delta B_{x}$ (see Section \ref{sec2} for details).} }
\label{fig2}
\end{figure}

\section{Overview and case studies \label{sec2}}

Figure \ref{fig1} presents \blue{an} overview of the considered 10 day interval. Over this interval the spacecraft was at radial distances from 0.17 to 0.24 au, that is from 35 to 50 solar radii, and remained within the inward magnetic field sector without crossing the heliospheric current sheet \citep{Szabo20:apj,Phan20:apjs}. No coronal mass ejections were observed in the considered interval \citep{Szabo20:apj,Phan20:apjs}. The spacecraft was approximately co-rotating with the Sun and connected to the same coronal hole \citep{Bale19:nature,badman20:apj}. Panels (a)--(d) present 1 min averages of magnetic field magnitude, proton flow velocity, densities and temperatures of protons and electrons as well as plasma density estimates provided by the quasi-thermal noise spectroscopy. The magnetic field magnitude was about 100 nT at the perihelion on November 6, and around 50 nT as the spacecraft moved to the radial distance of 50 solar radii from the Sun on November 1 and 10. The solar wind was typically slow with proton flow velocity below 400 km/s, except $\approx 20$ h period of fast solar wind between November 9 and 10. The three plasma density estimates in panel (c) are consistent with each other within a few tens of percent over the entire interval, except the fast solar wind interval, where proton densities were about four times smaller than plasma density estimates provided by the quasi-thermal noise spectroscopy. We will use the electron density estimates, calibrated to best match proton densities and the results of quasi-thermal noise spectroscopy, as plasma \blue{density estimates. The proton} temperature during the considered interval varied between about 10 to 100 eV, while \blue{the} electron temperature remained around 30 eV. Panel (e) presents 1 min averages of proton and \blue{electron} betas and shows that both quantities varied in the range from 0.1 to 5.

Panel (f) presents the percentage of magnetic field increments with PVI$>5$ computed for 1h intervals. The averaged percentage \blue{of 1.3\% is four} orders of magnitude larger than one would observe if the magnetic field increments had Gaussian probability distributions. \blue{The non-Gaussian distributions of magnetic field increments are consistent with previous observations in the solar wind \cite[e.g.,][]{Sorriso-Valvo99,Greco09:apjl,Chhiber21:apj}}. Panel (g) shows that the number of CSs per day varied from 500 to 1500, while the averaged occurrence rate is 1,120 CSs per day. The percentage of bifurcated CSs per day varies from 5 to 15\% with the averaged value around 10\%. Panel (h) shows that the number of CSs per hour \blue{varies from 0 to 200 with the averaged} value around 50 CSs per hour.

Figure \ref{fig2} presents several CSs from our dataset. \blue{The left panels present a non-bifurcated CS.} Panel (a) shows the magnetic field magnitude and three magnetic field components in the spacecraft coordinate system $XYZ$. The magnetic field rotates across the CS through shear angle $\Delta \theta\approx 77^{\circ}$. The magnetic field magnitude changes across the CS by $\Delta B\approx 8$ nT, while the mean of magnetic field magnitudes at the CS boundaries is $\langle B\rangle \approx 45.6$ nT. The magnetic field variation within CS is relatively large, the difference between maximum and minimum values of the magnetic field magnitude is $\Delta B_{max}\approx 25$ nT. Panel (b) presents the magnetic field in local CS coordinate system $\textbf{\emph{xyz}}$. The magnetic field $B_{x}$ varies across the CS by $\Delta B_{x}\approx 57.3$ nT. The CS is not perfectly symmetric, the mean of $B_{x}$ values at the CS boundaries is $\langle B_{x}\rangle \approx -6.8$ nT. The values of $B_{y}$ at the CS boundaries are similar and their mean value is $B_{g}\approx 30$ nT. The normal component $B_{z}$ is around zero at the CS boundaries and remains small within the CS. The CS is observed in a plasma with plasma density of 209 cm$^{-3}$, electron temperature of 25 eV, and proton temperature of 70 eV, so that electron and proton betas are $\beta_{e}\approx 1$ and $\beta_{p}\approx 2.8$. Using Eq. (\ref{eq:dbeta}) we found that plasma beta varies across the CS by $\Delta \beta\approx 1.4$.

Panels (c) and (d) present current densities $J_{x}$ and $J_{y}$ as well as components parallel and perpendicular to local magnetic field. The CS central region\blue{, where} $|B_{x}-\langle B_{x}\rangle|<0.2\Delta B_{x}$, is \blue{highlighted in panels (b)--(d)}. We characterize the CS intensity by $J_{peak}$, that is the absolute peak value of parallel current density $J_{||}$, and by $J_{0}$, that is the absolute value of parallel current density $J_{||}$ averaged over the CS central region. For the considered CS we have $J_{0}\approx 1.5\;{\rm \mu}$A/m$^{2}$ and $J_{peak}\approx 2.35\;{\rm \mu}$A/m$^{2}$ or $J_{0}\approx 0.63\; J_{A}$ and $J_{peak}\approx 1.05\;J_{A}$ in units of local Alfv\'en current density $J_{A}=eN_{0}V_{A}$, where $N_0$ is plasma density, $V_{A}=\langle B\rangle /(\mu_0 N_0 m_{p})^{1/2}$ is local Alfv\'en speed, $m_{p}$ and $e$ are proton mass and charge. The CS thickness is determined as follows
\begin{eqnarray}
\lambda= \frac{\Delta B_{x}}{2\mu_0 J_{0}}
\label{eq:thickness}
\end{eqnarray}
\blue{The CS} thickness is $\lambda\approx 14$ km, while in units of \blue{local} proton inertial length $\lambda_{p}$ and thermal proton gyroradius $\rho_{p}$ we have $\lambda\approx 0.9\lambda_{p}$ and $\lambda\approx 0.6\rho_{p}$. Note that strictly-speaking $\lambda$ is a half-thickness, because according to Eq. (\ref{eq:thickness}) the magnetic field can be approximated as $B_{x}\approx \langle B_{x}\rangle+0.5\Delta B_{x}\tanh(z/\lambda)$, but we keep to the terminology often used in theoretical studies and refer to this parameter as thickness.

The right panels in Figure \ref{fig2} present \blue{a bifurcated CS observed in a plasma with density of 343 cm$^{-3}$, electron temperature of 36 eV, and proton temperature of 15 eV, so that $\beta_{e}\approx 1.1$ and $\beta_{p}\approx 0.5$. The current densities $J_{y}$ and $J_{||}$ have bifurcated profiles and, accordingly, the magnetic field rotation occurs in two steps, in contrast to relatively smooth rotation in non-bifurcated CSs.} Since $J_{||}$ profile is bifurcated, the current density averaged over the CS central region does not reflect the actual CS intensity and cannot be used to estimate the CS thickness. We have determined the temporal duration of each bifurcated CS manually as the half of the temporal distance between the two steps of magnetic field rotation. The spatial scale corresponding to this temporal duration will be referred to as the thickness of a bifurcated CS. For the considered CS we have $\lambda\approx 36$ km \blue{that is around $3\lambda_{p}$ or $4.4\rho_{p}$}.
 
 


\section{Statistical properties\label{sec3}}

\begin{figure}[ht!]
\centering
\includegraphics[width=\textwidth]{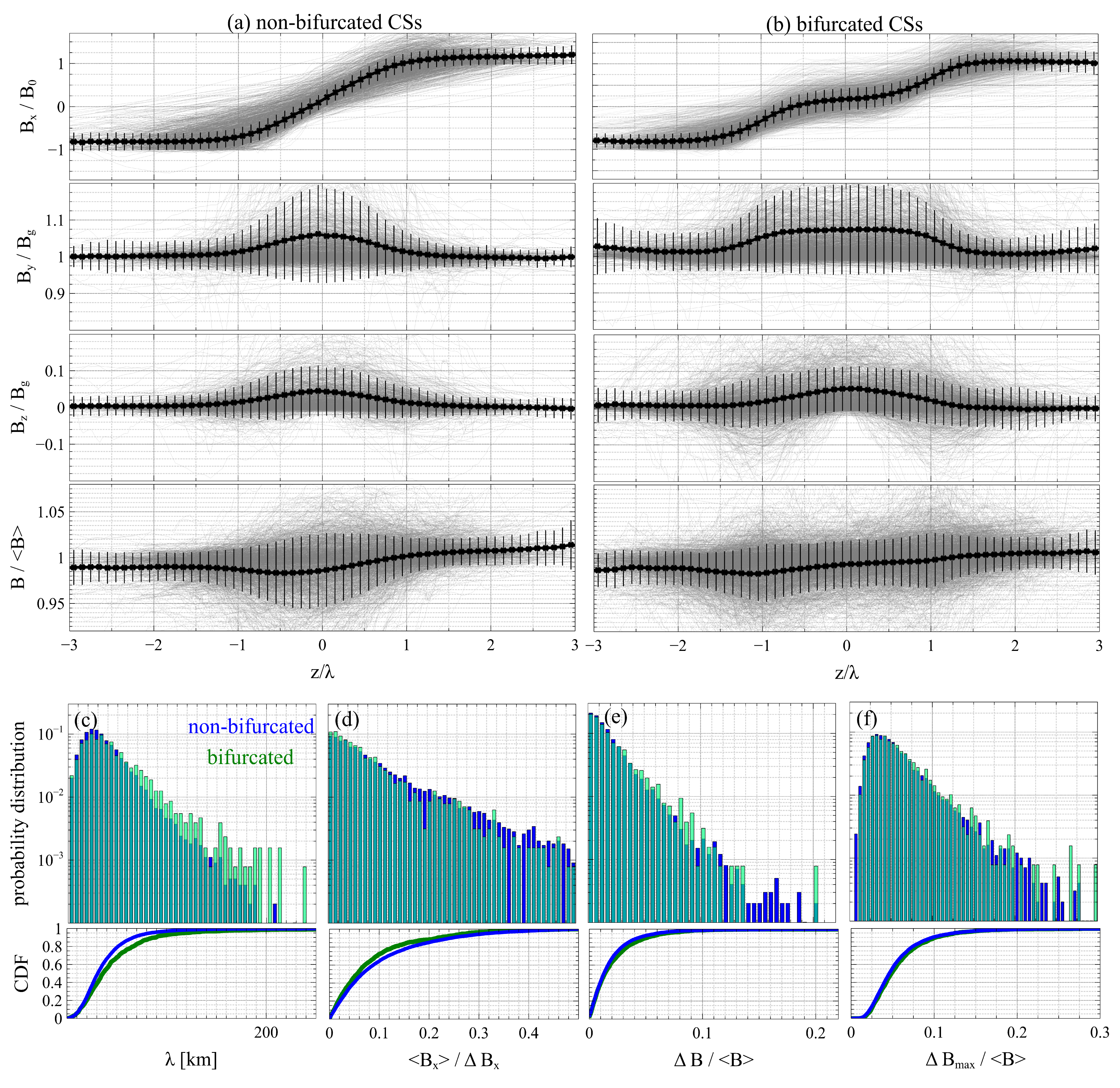}
\caption{Panels (a) and (b) present the epoch analysis of the non-bifurcated and bifurcated CSs (see Section \ref{sec3} for details). Individual CS profiles (gray) shown in panels (a) and (b) were appropriately normalized and aligned, the averaged profiles (black) are shown along with errors bars indicating the standard deviations. Individual $B_{x}$ profiles were normalized to $B_0$ that is the half difference of $B_{x}$ values at the right and left CS boundaries, individual $B_{y}$ and $B_{z}$ profiles were normalized to $B_{g}$ that is the half sum of $B_{y}$ values at the CS boundaries, individual $B$ profiles were normalized to $\langle B\rangle$ that is the half sum of $B$ values at the CS boundaries, the spatial coordinate $z$ across each CS was normalized to CS thickness $\lambda$. Panels (c)--(f) present statistical distributions of various parameters of bifurcated and non-bifurcated CSs including thickness $\lambda$, asymmetry $<B_x>/\Delta B_{x}$, relative variation $\Delta B/\langle B\rangle$ of the magnetic field magnitude across CS, and maximum relative variation $\Delta B_{max}/\langle B\rangle$ of the magnetic field magnitude within CS.}
\label{fig3}
\end{figure}

Figure \ref{fig3} presents averaged magnetic field profiles of non-bifurcated and bifurcated CSs. Before computing the averaged profiles, individual CS profiles were appropriately normalized and aligned. Individual profiles of $B_{x}$ and $B$ were respectively normalized to $B_0=0.5\Delta B_{x}$ and $\langle B\rangle$, where a signed quantity of $B_0$ was used so that $B_{x}/B_0$ was always negative/positive at the left/right boundary. Individual profiles of $B_{y}$ and $B_{z}$ were both normalized to $B_{g}$. The individual profiles were aligned by normalizing the spatial distance $z$ to CS thickness $\lambda$ and setting $z=0$ at the CS center. The center of a non-bifurcated CS corresponds to $B_{x}=\langle B_{x}\rangle$, while for a bifurcated CS it is in the middle between \blue{the two steps of magnetic field rotation}. Each $B_{x}/B_0$ profile with smaller absolute value at the right boundary was reflected with respect to $z=0$ and multiplied by $-1$. The other magnetic field profiles corresponding to that $B_{x}/B_0$ profile were reflected too. The reflection procedure allows us to keep the smaller of the $B_{x}/B_0$ absolute values at the left boundary, so that the averaged CS asymmetry can be revealed. 

The averaged profiles of bifurcated and non-bifurcated CSs along with individual CS profiles are shown in panels (a) and (b) of Figure \ref{fig3}. The averaged $B_{x}/B_0$ profiles demonstrate that both bifurcated and non-bifurcated CSs are typically asymmetric with left and right boundary values around $-0.75$ and 1.25, respectively. In addition, the averaged $B_{x}/B_0$ profiles show that magnetic field rotation occurs relatively smoothly in non-bifurcated CSs and, in contrast, in two steps in bifurcated CSs. The averaged $B_{y}/B_{g}$ profiles demonstrate that statistically this magnetic field component has similar values at the CS boundaries and a few percent larger value around the CS central region. The averaged $B_{z}/B_{g}$ profiles show that the normal component is around zero \blue{at the CS boundaries and remains} small within CS. Note that each individual $B_{z}/B_{g}$ profile was multiplied by its sign around the CS central region to reveal the absolute value of $B_{z}/B_{g}$ in the averaged profile. The averaged profiles of $B/\langle B\rangle $ show that the magnetic field magnitude varies within both CS types by only a few percent. The magnetic field magnitude is larger at the right boundary that is consistent with the asymmetry of $B_{x}$ profiles. In other words, the averaged profiles indicate that $\Delta (B^2)\approx \Delta (B_{x}^2+B_{y}^{2})\approx \Delta (B_{x}^{2})$, which can be rewritten as follows
\begin{eqnarray}
\langle B\rangle \Delta B\approx \langle B_{x}\rangle\Delta B_{x},
\label{eq:dBvsdBx}
\end{eqnarray}
\blue{where we took into account that $\Delta (B^2)=2\langle B\rangle \Delta B$ and $\Delta (B_{x}^{2})=2\langle B_{x}\rangle \Delta B_{x}$.}

\begin{figure}[ht!]
\centering
\includegraphics[width=0.5\textwidth]{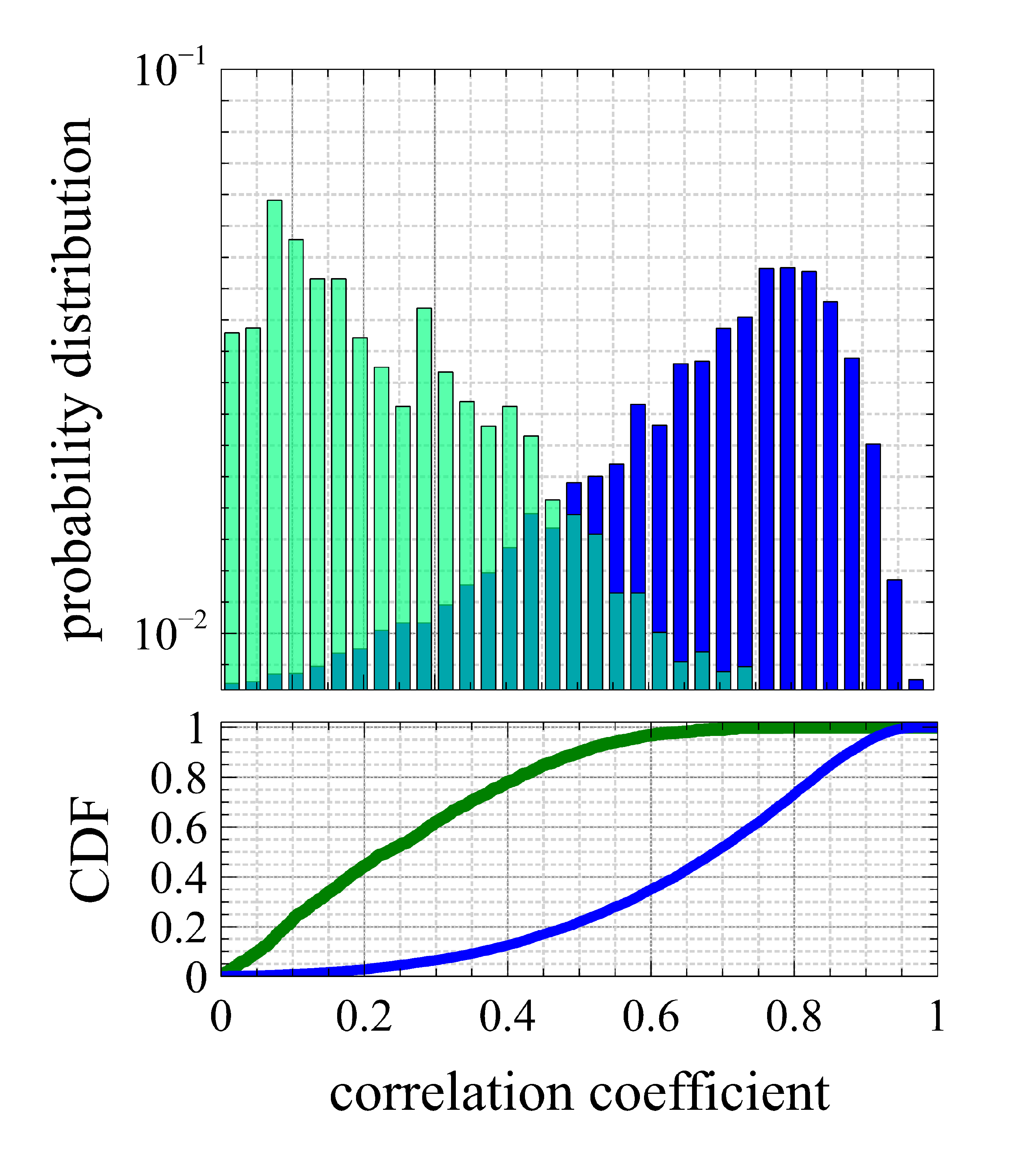}
\caption{Panel (a) shows the probability distributions of the correlation coefficient between parallel current density profile $J_{||}$ and a model non-bifurcated profile $\langle J_{||}\rangle\;{\rm sech}^{2}(V_{n}t/\lambda)$, where $\langle J_{||}\rangle$ is the parallel current density averaged over the CS central region, $t=0$ corresponds to the CS center, $V_{n}$ is the normal component of proton flow velocity and $\lambda$ is CS thickness determined by Eq. (\ref{eq:thickness}). Panel (b) presents the corresponding cumulative distribution functions. Both panels demonstrate the adequacy of our visual classification of bifurcated and non-bifurcated CSs.}
\label{fig4}
\end{figure}

The statistical distributions in panels (c)--(f) of Figure \ref{fig3} show that bifurcated and non-bifurcated CSs have similar distributions of the major CS parameters. The thickness of both CS types is in the range from a few to 200 km with the typical value around 30 km. Both CS types are typically asymmetric, $\langle B_{x}\rangle /\Delta B_{x}\gtrsim 0.1$ for \blue{about} 40\% of the CSs, with relatively small variation of the magnetic field magnitude within \blue{CS and between CS boundaries, $\Delta B_{max}/\langle B\rangle \lesssim 0.1$ and $\Delta B/\langle B\rangle \lesssim 0.1$} for about 95\% of the CSs. The only substantial difference between the two CS types is that magnetic field rotation occurs smoothly within non-bifurcated CSs and in two steps within bifurcated CSs. This difference can be also demonstrated quantitatively. For each CS we compute \blue{a} correlation coefficient between $J_{||}$ profile and a model non-bifurcated profile $\langle J_{||}\rangle\; {\rm sech}^{2}(V_{n}t/\lambda)$, where $V_{n}$ is \blue{the normal component of local} proton flow velocity, $t=0$ corresponds to $B_{x}=\langle B_{x}\rangle$, and $\lambda$ was determined by Eq. (\ref{eq:thickness}) for all CSs. The probability distributions in Figure \ref{fig4} demonstrate that the correlation coefficient is below (above) 0.5 for more than 90\% (80\%) of the bifurcated (non-bifurcated) CSs, which proves the adequacy of our visual \blue{CS} classification. 

\begin{figure}[ht!]
\centering
\includegraphics[width=\textwidth]{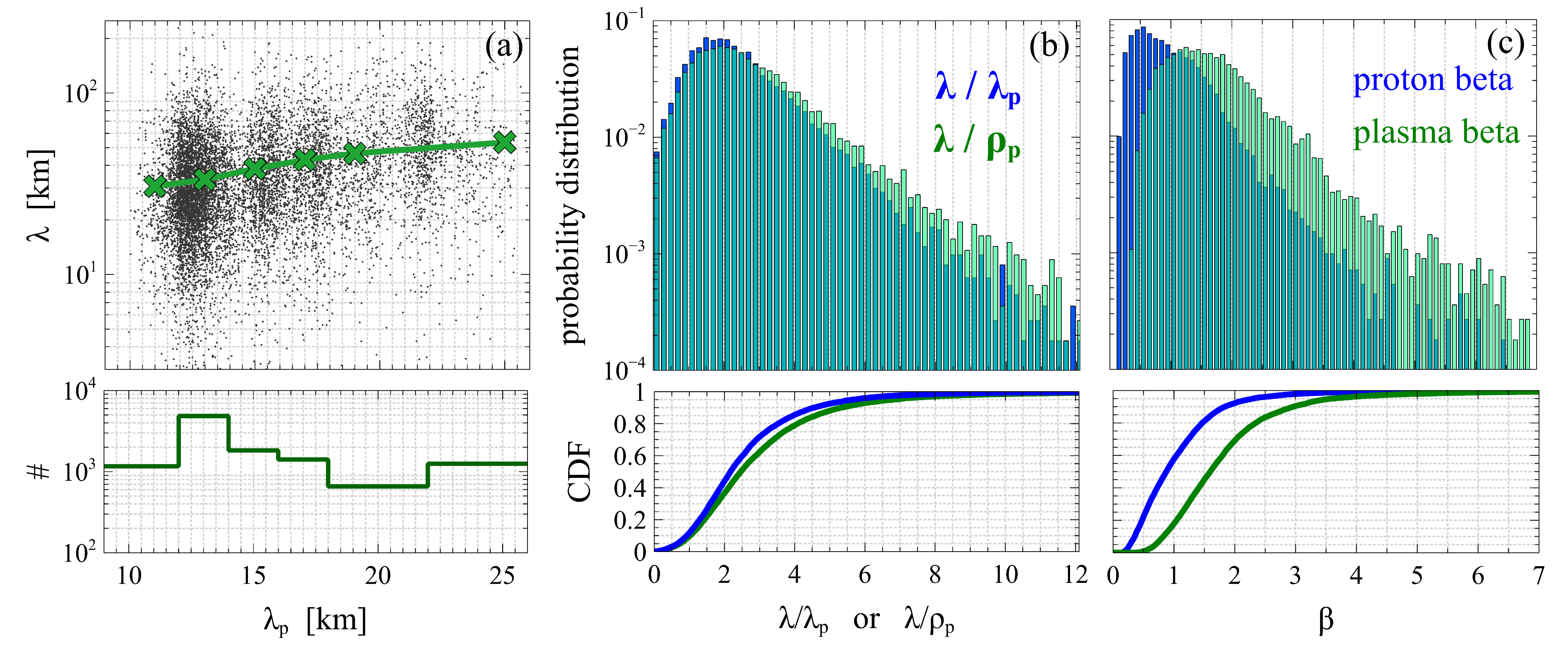}
\caption{Panel (a) presents a scatter plot of CS thickness $\lambda$ versus proton inertial length $\lambda_p$. The green curve represents bin averaged values of CS thickness with the number of CSs within each bin shown at the bottom. Panel (b) shows the statistical distributions of CS thickness in units of proton inertial length $\lambda_{p}$ and thermal proton gyroradius $\rho_{p}$. Panel (c) presents statistical distributions of proton and total plasma betas for the considered CSs.}
\label{fig5}
\end{figure}

Figure \ref{fig5} compares the thickness of the CSs to local proton inertial length $\lambda_{p}$ and thermal proton gyroradius $\rho_{p}=\lambda_{p}\beta_{p}^{1/2}$, where $\beta_{p}$ is proton beta. Panel (a) shows that although local proton inertial length for the CSs varied only between 10 and 25 km, there is a trend that the CSs observed at larger $\lambda_{p}$ tend to have larger thickness. Panel (b) presents the probability distribution of $\lambda/\lambda_{p}$ and shows that the CS thicknesses are in the range from about 0.1 to 10$\lambda_{p}$ with the typical value around $2\lambda_{p}$. Thus, the collected CSs are structures at proton kinetic scales, with about 10\% of the CSs at sub-proton scales, $\lambda\lesssim \lambda_{p}$. Panel (b) shows that the probability distribution of $\lambda/\rho_{p}$ is identical with the one of $\lambda/\lambda_{p}$, because according to panel (c) proton beta $\beta_{p}$ is between 0.4 and 2 for more than 80\% of the CSs. The total plasma beta  was between 0.5 and 3 for about 90\% of the CSs with the typical value around 1.5.

\section{Tests of the diamagnetic suppression condition\label{sec4}}

\citet{Swisdak_2010} showed that magnetic reconnection in a planar CS with magnetic shear angle $\Delta \theta$ and plasma beta variation $\Delta \beta$ between the CS boundaries is allowed/suppressed if the following condition is satisfied/violated \citep[see also][]{Swisdak03:jgr}
\begin{eqnarray}
\Delta \beta\lesssim 2(L/\lambda_{p})\tan(\Delta \theta/2),
\label{eq:rec_condition}
\end{eqnarray}
where $L$ is the scale of plasma pressure gradient across X-line, which should be on the order of one for magnetic reconnection to be sufficiently fast \citep[e.g.,][]{Cassak06:apj}. The violation of this criterion results in suppression of magnetic reconnection, because the diamagnetic drift of X-line becomes comparable with the characteristic Alfv\'{e}n speed \citep{Swisdak_2010}. \blue{Note that condition (\ref{eq:rec_condition}) is necessary, but not sufficient for magnetic reconnection}. Although the magnetic field magnitude does not substantially vary across the CSs (Figure \ref{fig3}e), it does result in plasma beta variation $\Delta \beta$, which might be of importance for reconnection development. Before testing the diamagnetic suppression condition (\ref{eq:rec_condition}), we \blue{address} the origin of the magnetic field magnitude variation across \blue{the CSs}.

\begin{figure}[ht!]
\centering
\includegraphics[width=\textwidth]{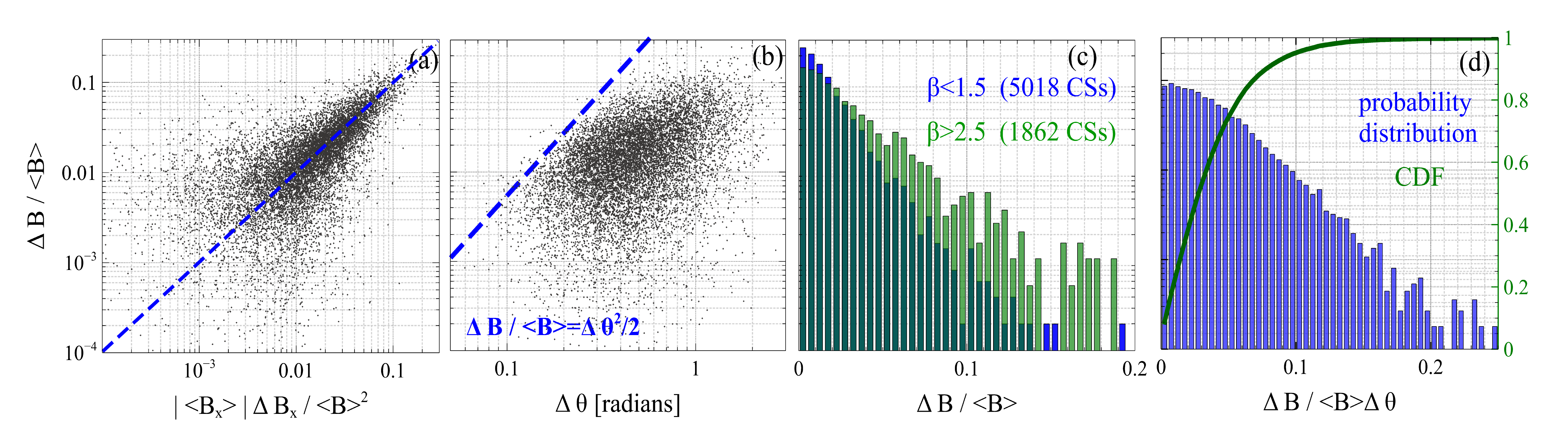}
\caption{The scatter plots of (a) $\Delta B/\langle B\rangle$ versus $|\langle B_x\rangle|\Delta B_x/\langle B\rangle^2 $ and (b) $\Delta B/\langle B\rangle$ versus shear angle $\Delta \theta$. Panel (c) presents the probability distributions of $\Delta B/\langle B\rangle$ for the CSs observed at various plasma betas, $\beta<1.5$ and $\beta>2.5$. Panel (d) presents the probability distribution and corresponding cumulative distribution function of $\Delta B/\langle B\rangle \Delta \theta$ for all the CSs in our dataset.}
\label{fig6}
\end{figure}

Figure \ref{fig6} presents a further experimental test of Eq. (\ref{eq:dBvsdBx}) and clarifies the dependence of $\Delta B/\langle B\rangle$ on local plasma parameters. Panel (a) shows there is a correlation between $\Delta B/\langle B\rangle$ and $\langle B_{x}\rangle \Delta B_{x}/\langle B\rangle^2$, especially at $\Delta B/\langle B\rangle\gtrsim 0.01$. This confirms that the magnetic field variation across the CSs is predominantly due to the $B_{x}$ asymmetry. Panel (b) shows there is a positive correlation between $\Delta B/\langle B\rangle$  and $\Delta \theta$ and reveals the upper threshold, $\Delta B/\langle B\rangle\lesssim \Delta \theta^2/2$. This threshold follows from Eq. (\ref{eq:dBvsdBx}) once we take into account that $\langle B_{x}\rangle\lesssim 0.5\Delta B_{x}$ and $\Delta B_{x}\approx \langle B\rangle \Delta \theta$. Panel (c) presents the probability distribution functions of $\Delta B/\langle B\rangle$ for the CSs observed at $\beta<1.5$ and $\beta>2.5$. These distributions show that larger values of $\Delta B/\langle B\rangle$ are observed at higher plasma betas. Panel (d) shows the probability distribution of $\Delta B/\langle B\rangle\Delta \theta$, which is a proxy of the ratio between perpendicular and parallel current densities (see Eq. (\ref{eq:Jper_Jpar})). For more than 95\% of the CSs we have $\Delta B/\langle B\rangle\Delta \theta\lesssim 0.1$, so that the current density in the CSs is dominated by the parallel component. Similarly we have found that statistically $\Delta B_{max}/\langle B\rangle\ll \Delta \theta$ (not show here). Note that similarly to $\Delta B/\langle B\rangle$ we observe larger values of  $\Delta B/\langle B\rangle\Delta \theta$ at larger plasma betas (not shown here).

\begin{figure}[ht!]
\centering
\includegraphics[width=\textwidth]{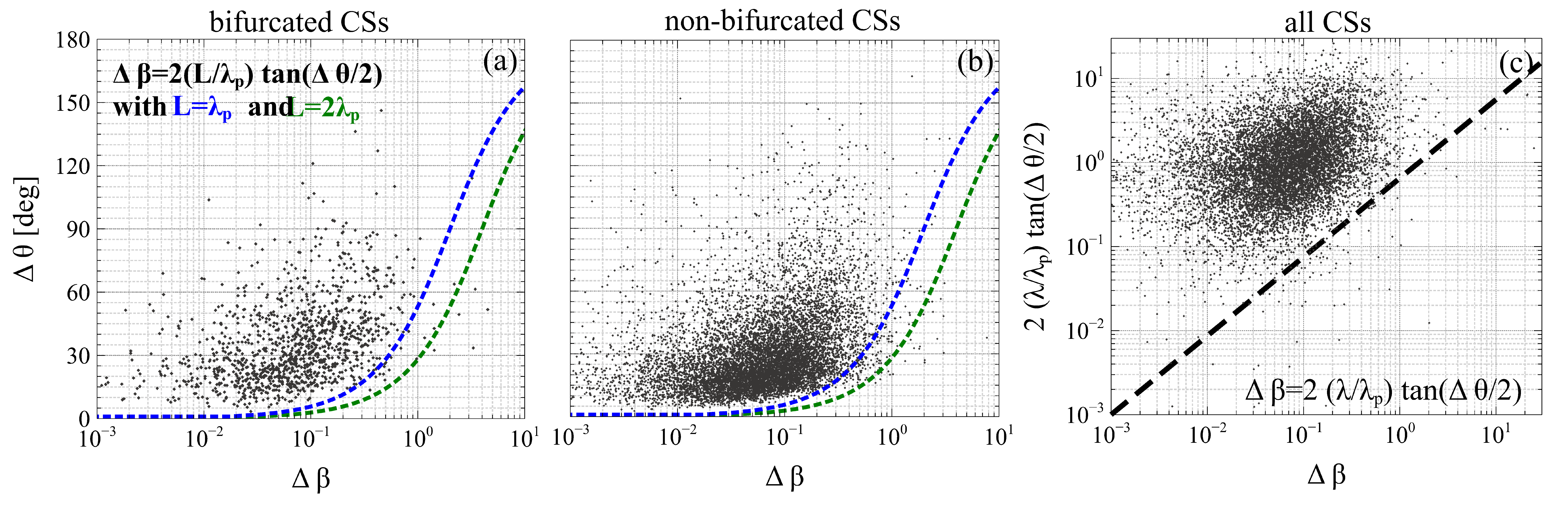}
\caption{The test of the diamagnetic suppression condition (\ref{eq:rec_condition}) for the considered 11,200 CSs. Panels (a) and (b) present the test of this condition with $L=\lambda_{p}$ and $2\lambda_{p}$ separately for bifurcated and non-bifurcated CSs, while panel (c) presents the test of this condition with $L=\lambda$. }
\label{fig7}
\end{figure}

Figure \ref{fig7} presents a test of the diamagnetic suppression condition (\ref{eq:rec_condition}). Panels (a) and (b) show that all bifurcated as well as non-bifurcated CSs satisfy condition (\ref{eq:rec_condition}) with $L=2\lambda_{p}$ and only about 1\% of the CSs of both types violate this condition with $L=\lambda_{p}$. In \cite{Swisdak_2010} theory, parameter $L$ is the scale of the plasma pressure gradient across X-line, while for each CS we could estimate only local CS thickness $\lambda$. \blue{It is reasonable to test} condition (\ref{eq:rec_condition}) for $L=\lambda$, because $\lambda$ characterizes the local scale of plasma pressure gradient across a CS before reconnection potentially occurs. Panel (c) shows that about 99\% of the CSs satisfy condition (\ref{eq:rec_condition}) with $L=\lambda$. \blue{Since the previous observations at 3s temporal or, equivalently, $\approx 1000$ km spatial resolution showed that only a relatively small fraction of the CSs in the solar wind are reconnecting \citep[e.g.,][]{Gosling12,Osman14:prl}}, the fact that condition (\ref{eq:rec_condition}) is satisfied by almost all the CSs implies that the diamagnetic suppression condition does not control reconnection onset and occurrence in the solar wind.

\section{Scale-dependent CS properties and critical current density\label{sec5}}

\begin{figure}[ht!]
\centering
\includegraphics[width=\textwidth]{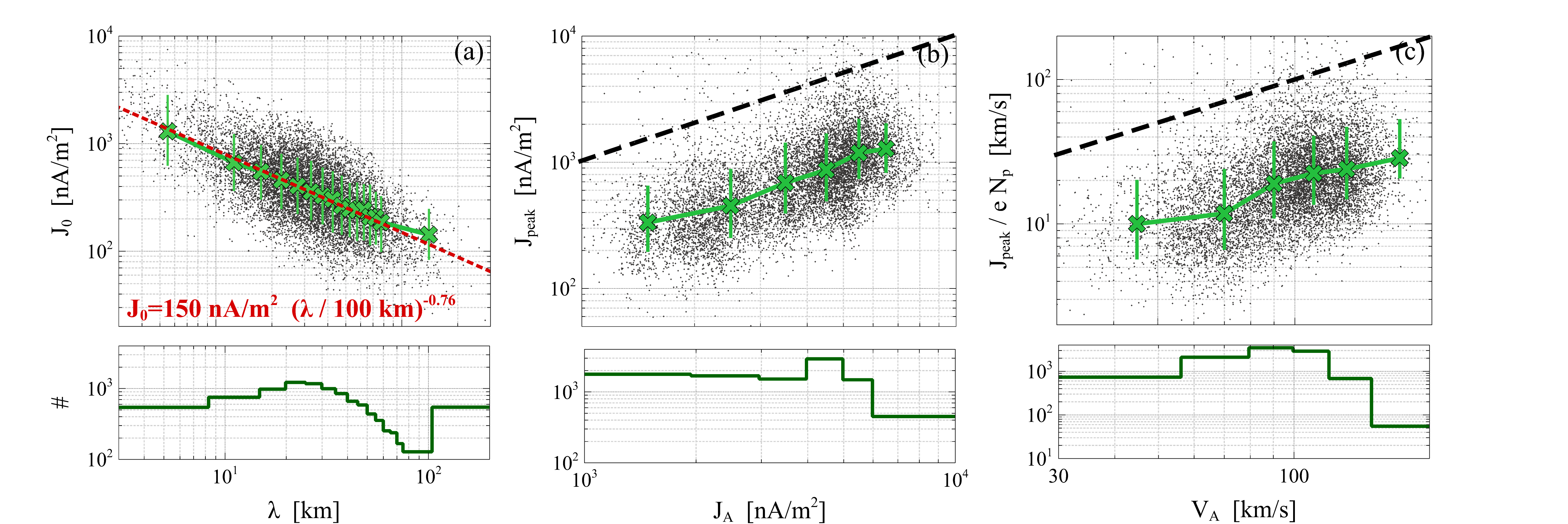}
\caption{The scatter plots of (a) current densities $J_0$ versus CS thicknesses $\lambda$, (b) peak current densities $J_{peak}$ versus local Alfv\'{e}n current densities $J_{A}=eN_0V_{A}$, and (c) peak values of the ion-electron drift velocity $J_{peak}/eN_0$ versus local Alfv\'{e}n speed $V_A$. The black dashed line in panels (b) and (c) corresponds to $J_{peak}=J_{A}$. In each panel a green curve represents median values of the respective quantity within the bins shown at the bottom of each panel, where also the number of CSs within each bin is presented. The errors bars of a green curve correspond to 15th and 85th percentiles of the corresponding quantity within each bin. The best power-law fit of the scattered data (red curve) is indicated in panel (a) along with the best fit parameters.}
\label{fig8}
\end{figure}

Figure \ref{fig8} presents the analysis of \blue{averaged and absolute peak values, $J_0$ and $J_{peak}$, of the parallel current density within the CSs. In this} section we do not consider bifurcated CSs (about $10\%$ of all CSs), because $J_0$ does not reflect their actual current density. Panel (a) shows that the current density $J_0$ is inversely correlated with CS thickness $\lambda$, so that smaller-scale CSs tend to be more intense. We reveal the trend, shown in panel (a), by sorting the CSs into bins corresponding to different values of $\lambda$, and computing the median value of $J_0$ within each bin. The error \blue{bars in panel (a)} correspond to 15th and 85th percentiles of $J_0$ within each bin. The number of the CSs within each bin is shown at the bottom of panel (a). We also fitted the scattered data in panel (a) by a power-law function and found the following best fit
\begin{eqnarray}
J_{0}=150\;{\rm nA\;m^{-2}}\cdot \left(\frac{\lambda}{100\;{\rm km}}\right)^{-0.76},
\label{eq:J_scale}
\end{eqnarray}
which well describes the median profile revealed by binning the data. The median profiles in panels (b) and (c) show that the current density $J_{peak}$ and corresponding ion-electron drift velocity $J_{peak}/eN_{0}$ are positively correlated with local Alfv\'{e}n current density $J_{A}=eN_{0}V_{A}$ and Alfv\'{e}n speed $V_{A}$, respectively. The CSs tend to be more intense \blue{at larger} Alfv\'{e}n speed. In addition, panels (b) and (c) show that the current densities $J_{peak}$ are statistically below some threshold, $J_{peak}\lesssim J_{A}$ for more than 99\% of the CSs and $J_{peak}\lesssim J_{A}/2$ for about 98\% of the CSs.

\begin{figure}[ht!]
\centering
\includegraphics[width=0.6\textwidth]{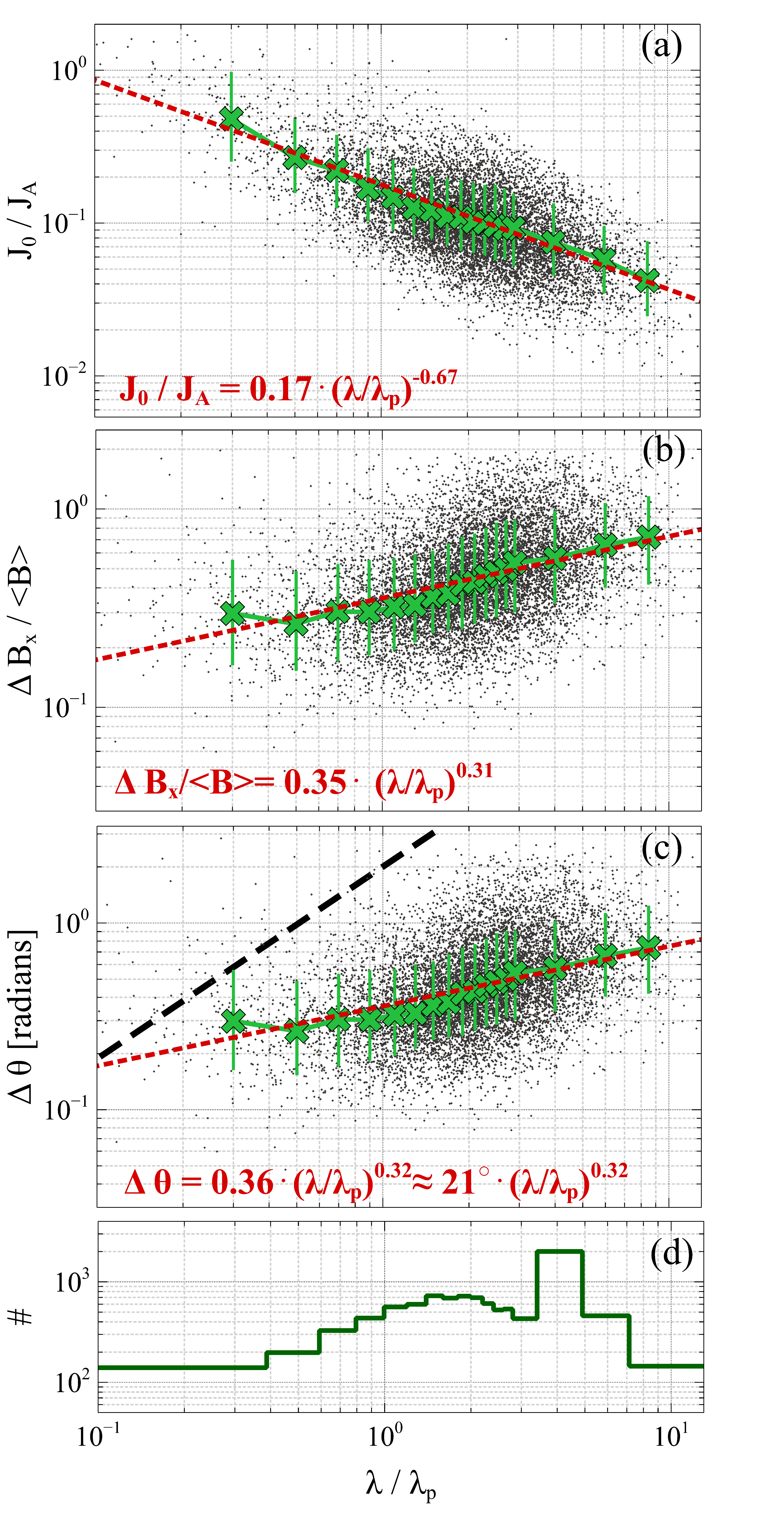}
\caption{The scatter plots of (a) current densities normalized to local Alfv\'{e}n current density $J_0/J_{A}$ versus normalized CS thicknesses $\lambda/\lambda_p$; (b) normalized CS amplitudes $\Delta B_{x}/\langle B\rangle$ versus $\lambda/\lambda_p$; (c) shear angles $\Delta \theta$ versus $\lambda/\lambda_p$. The green curves in the panels represent median values of the respective quantities within the bins shown in panel (d) along with the number of CSs within each bin. The errors bars of the median profiles correspond to 15th and 85th percentiles of the quantities within the bins. The best power-law fits (red curves) of the scattered data are shown in the panels along with the best fit parameters.}
\label{fig9}
\end{figure}

Figure \ref{fig9} reveals scale-dependence \blue{of various CS properties in normalized units}. The scale-dependent trends of the various quantities in panels (a)--(c) \blue{are demonstrated by  the median profiles revealed by sorting the CSs into bins. The number of CSs within each bin} is shown in panel (d). Panel (a) \blue{shows that}  normalized current density, $J_0/J_{A}$, is inversely correlated with normalized CS thickness, $\lambda/\lambda_{p}$. The least squares fitting of the scattered data by a power-law function reveals the following best fit
\begin{equation} \label{eq:CS_internsity_scale_dependence}
    J_0/J_A = 0.17\;(\lambda/\lambda_p)^{-0.67},
    \label{eq:JJA_scale}
\end{equation}
which also well describes the median profile \blue{in} panel (a). Panel (b) shows that the CS amplitude $\Delta B_x/\langle B\rangle$ is positively correlated with $\lambda/\lambda_p$. The \blue{best fit by} a power-law function
\begin{equation} \label{eq:CS_amplitude_scale_dependence}
    \Delta B_x/\langle B\rangle = 0.35\;(\lambda/\lambda_p)^{0.31},
        \label{eq:BxB_scale}
\end{equation} 
again well describes the median profile in panel (b). Finally, since in a CS with relatively constant magnetic field magnitude, the CS amplitude $\Delta B_{x}/\langle B \rangle$ is proportional to shear angle $\Delta \theta$, the latter is expected to be scale-dependent. Panel (c) confirms the scale-dependency of the magnetic field shear angle. \blue{The best fit by a power-law function} 
\begin{equation} \label{eq:CS_shearangle_scale_dependence}
    \Delta \theta = 0.36\;(\lambda/\lambda_p)^{0.32}\approx  21^{\circ}\;(\lambda/\lambda_p)^{0.32},
    \label{eq:teta_scale}
\end{equation}
well describes the median profile in panel (c).

\section{Discussion \label{sec6}}

In this paper we presented a statistical analysis of 11,200 CSs in the near-Sun solar wind observed aboard PSP \blue{around 0.2 au, at the} radial distances of 35--50 solar radii from the Sun. \blue{These CSs} are proton kinetic-scale structures with thickness in the range from about 0.1 to 10$\lambda_{p}$ with the typical value around $2\lambda_{p}$, where $\lambda_{p}$ is local proton inertial length. The CSs have similar scales in units of local thermal proton gyroradius, because proton beta was around one (Section \ref{sec1}). The resolution of these thin CSs became possible thanks to high resolution (73--290 S/s) magnetic field measurements provided by the FIELDS instrument suite \citep{Bale16:ssr,Bowen20:jgr}. The previous studies of CSs in the near-Sun solar wind \blue{were} limited to magnetic field measurements at 0.2s resolution aboard Helios \citep{soding01} and PSP \citep{Phan20:apjs} spacecraft, so that only CSs with thickness larger than 10$\lambda_{p}$ were resolved. \blue{Note that our CS dataset is biased toward the thinnest resolvable CSs, because we focused on proton kinetic-scale CSs, which are expected to be crucial for turbulence dissipation and development of turbulence cascade \citep[e.g.,][]{Servidio15:jpp,Cerri&Califano17,Franci17,Franci18,Papini19:apj}. This dataset is sufficiently representative though to address the properties and origin of proton kinetic-scale CSs in the near-Sun solar wind.} In this section we summarize the results of this study and compare the properties of kinetic-scale CSs in the near-Sun solar wind to those reported at 1 au \citep{Vasquez07,Vasko21a:apjl,Vasko21b:apjl}.

We have found the occurrence rate of proton kinetic-scale CSs \blue{around 0.2 au to be} on average 1,120 CS/day, while at 1 au it is about 150 CS/day \citep{Vasquez07,Vasko21a:apjl}. \blue{The radial trend of the occurrence rate consistent with observations at 0.2 and 1 au would be $\approx 1/R^{1.3}$, where $R$ is the radial distance from the Sun.} The larger occurrence rate closer to the Sun is consistent with the previous studies of larger-scale CSs at the radial distances of 0.3--19 au \citep{tsurutani79,soding01}. We stress however that the actual trend can be affected by the selection procedure of the CSs. For example, in this study as well as at 1 au only PVI index computed at the minimum time increment was used to collect the CSs. The time increment was 1/73--1/290s in this study and 1/3--1/11s at 1 au \citep{Vasquez07,Vasko21a:apjl,Vasko21b:apjl}.

We have found that the CSs in the near-Sun solar wind are typically asymmetric with small relative variations of the magnetic field magnitude \blue{between CS boundaries as well as} within CS (Section \ref{sec3}). \blue{We observe that these variations,} $\Delta B/\langle B\rangle$ and $\Delta B_{max}/\langle B\rangle$, are statistically much smaller than shear angle $\Delta \theta$, so that the magnetic field variation within the CSs is predominantly rotation, rather than variation of the magnetic field magnitude. Accordingly, the current density in the CSs is dominated by the magnetic field-aligned component (Section \ref{sec4}). We have found that about 10\% of the CSs are bifurcated that is magnetic field rotation within CS occurs in two steps, which is often seen in reconnecting CSs \citep[e.g.,][]{Gosling&Szabo08,Phan10,Mistry17:jgr}. Note though that the fraction of reconnecting CSs in our dataset could not be determined, because the cadence of proton measurements of 0.2--0.9s is insufficient to establish the presence or absence of plasma jets at proton kinetic scales, while a bifurcated magnetic field profile does not necessarily imply reconnection \citep{Gosling&Szabo08,Schindler08:phpl}. All the CS properties (asymmetry, small variations of the magnetic field magnitude, \blue{dominance of parallel current}, $\approx 10\%$ of bifurcated CSs) in the near-Sun solar wind are \blue{identical with those at 1 au} \citep{Vasquez07,Vasko21a:apjl,Vasko21b:apjl}. \blue{Moreover,} the averaged magnetic field profiles of the CSs in the near-Sun solar wind (Figure \ref{fig3}) are \blue{identical with those at 1 au} \citep{Vasko21a:apjl}.


We have found that the relative variation $\Delta B/\langle B\rangle$ of the magnetic field magnitude across the CSs is statistically small, but tends to be larger at larger plasma betas (Section \ref{sec4}). The same dependence of $\Delta B/\langle B\rangle$ on $\beta$ has been observed at 1 au \citep{Vasko21a:apjl}. The magnetic field magnitude variation translates into plasma beta variation $\Delta \beta$, which could, in principle, control magnetic reconnection via the diamagnetic suppression mechanism \citep{Swisdak03:jgr,Swisdak_2010}. We have shown that condition (\ref{eq:rec_condition}) necessary for reconnection to occur is satisfied by almost all of the CSs. \blue{Since only a small fraction of the CSs in the solar wind are expected to be reconnecting \citep[e.g.,][]{Gosling12,Osman14:prl}, this implies that the presence or absence of reconnection within proton kinetic-scale CSs are not controlled} by the diamagnetic mechanism. \blue{The studies at 1 au} showed that condition (\ref{eq:rec_condition}) is satisfied \blue{not only} by reconnecting CSs \citep{Phan10,Gosling&Phan13}, \blue{but} by almost all proton kinetic-scale CSs too \citep{Vasko21a:apjl}. Moreover, \cite{Vasko21a:apjl} have shown that this condition is satisfied automatically due to the geometry of solar wind CSs dictated by their source, which is turbulence cascade according to the previous studies at 1 au \citep{Vasquez07,Greco08,Greco09:apjl,Perri12:prl,zhdankin12,Vasko21b:apjl}. Thus, similarly to CSs at 1 au, the proton kinetic-scale CSs in the near-Sun solar wind automatically satisfy the necessary condition for reconnection and since reconnection likely occurs only in a small fraction of the CSs \citep{Gosling12,Phan20:apjs}, we conclude that the diamagnetic suppression condition does not control magnetic reconnection in the solar wind.

We have found that the CSs with smaller thickness tend to have larger averaged current density, $J_{0}\approx 150\;{\rm nA\;m^{-2}}\cdot  ( \lambda\;/\;100\;{\rm km})^{-0.76}$. The peak current density values $J_{peak}$ are in the range from 100 nA/m$^{2}$ to 10 ${\rm \mu}$A/m$^{2}$  (Figure \ref{fig8}), with the typical value around 700 nA/m$^{2}$. The current densities and corresponding ion-electron drift velocities tend to be larger at larger local \blue{Alfv\'{e}n} speed $V_{A}$ and Alfv\'{e}n current density $J_{A}$ (Figure \ref{fig8}). The trends observed in Figure \ref{fig8} for the CSs in the near-Sun solar wind are equivalent to those observed at 1 au, where it was found that $J_{0}\approx 6\;{\rm nA\;m^{-2}}\cdot  (\lambda\;/\;100\;{\rm km})^{-0.51}$, and $J_{peak}$ and $J_{peak}/eN_{0}$ are positively correlated with $J_{A}$ and $V_{A}$ \citep{Vasko21b:apjl}. The most probable value of the current density at 1 au is around 5 nA/m$^{2}$ \citep{Podesta17:jgr,Vasko21b:apjl}. The typical current densities \blue{observed at 0.2 and 1 au imply the} radial trend, $J_{peak}\propto R^{-3.1}$, that is in accordance with the fact that $J_{peak}\propto J_{A}\propto \left(N_{0}\right)^{1/2}\langle B\rangle\propto R^{-3}$, where we assumed $N_{0}\propto R^{-2}$ and $\langle B\rangle\propto R^{-2}$. Thus, the different properties of the CSs at 0.2 and 1 au in physical units are due to different background plasma density and magnetic field. In normalized units, proton kinetic-scale CSs at 0.2 and 1 au have similar current densities, $J_0/J_{A}$ and $J_{peak}/J_{A}$.

We have revealed remarkable scale-dependencies of \blue{normalized} current density $J_{0}/J_{A}$, magnetic field amplitude $\Delta B_{x}/\langle B\rangle$ and magnetic shear angle $\Delta \theta$ on normalized CS thickness $\lambda/\lambda_{p}$. Similar scale-dependencies have been recently reported at 1 \blue{au} \citep{Vasko21b:apjl}:
\begin{eqnarray}
J_{0}/J_{A}&=&0.17\;\cdot \left(\lambda/\lambda_{p}\right)^{-0.51}\nonumber\\
\Delta B_{x}/\langle B\rangle&=&0.33\;\cdot \left(\lambda/\lambda_{p}\right)^{0.49}\\
\Delta \theta&\approx& 19^{\circ}\;\cdot\left(\lambda/\lambda_{p}\right)^{0.5}\nonumber
\label{eq:one_AU}
\end{eqnarray}
The comparison of these trends to Eqs. (\ref{eq:JJA_scale})--(\ref{eq:teta_scale}) shows that only power-law indexes are slightly different, while the typical values of \blue{normalized CS properties} are similar. This substantiates that the CSs at 0.2 and 1 \blue{au} have similar properties in normalized units. \cite{Vasko21b:apjl} interpreted the scale-dependencies as a strong argument in support of the hypothesis that \blue{kinetic-scale} CSs in the solar wind are produced by turbulence \blue{cascade. The observation of similar scale-dependencies in the near-Sun solar wind indicates that proton kinetic-scale CSs in that region are also produced by turbulence cascade.}


\blue{Before summarizing the results, we note that electron densities were used as the plasma density estimates in those study. The electron densities were consistent within 10\% with ion density estimates and results of quasi-thermal noise spectroscopy in the entire 10 day interval, except for about 20 hours on November 9 and 10 (Figure \ref{fig1}). We have checked however that the use of ion density estimates or plasma density estimates by quasi-thermal noise spectroscopy does not affect any conclusions of this study.}




\section{Conclusions\label{sec7}}

We have analyzed 11,200 proton kinetic-scale current sheets observed by Parker Solar Probe in the near-Sun solar wind during 10 days around the first perihelion. The results of this study can be summarized as follows
\begin{enumerate}
    \item The CSs have thickness from a few to about 200 km with the typical value around 30 km. In terms of proton kinetic scales, the thickness of the CSs is from about 0.1 to 10 proton inertial lengths or thermal proton gyroradii. The current densities of the CSs are in the range from 100 nA/m$^{2}$ to 10 ${\rm \mu}$A/m$^{2}$ with the typical value around 700 nA/m$^{2}$. About 10\% of the CSs are bifurcated with magnetic field rotation within CS occurring in two steps, in contrast to relatively smooth rotation in non-bifurcated CSs. The properties of bifurcated and non-bifurcated CSs are essentially identical. \\

    \item  The magnetic field magnitude does not substantially vary within CS and, accordingly, the current density is dominated by the magnetic field-aligned component. Nevertheless, the CSs are typically asymmetric; that is, the magnetic field component $B_{x}$ reversing sign across CS has different absolute values at the CS boundaries and, accordingly, the magnetic field magnitudes at the CS boundaries are statistically different. This asymmetry results in plasma beta variation across the CSs.
    
    \item The analysis of plasma beta variations and magnetic shear angles across the CSs showed that magnetic reconnection within proton kinetic-scale current sheets is \blue{not controlled} by the diamagnetic suppression \blue{mechanism}. 
    
    \item The CSs with smaller thickness tend to have larger current densities and the best power-law fit of the trend is given by Eq. (\ref{eq:J_scale}). The current densities and corresponding ion-electron drift velocities are larger at larger local Alfv\'{e}n current densities $J_{A}$ and Alfv\'{e}n speeds $V_{A}$.
    
    \item The normalized quantities such as amplitudes $\Delta B_{x}/\langle B\rangle $, current densities $J_0/J_{A}$ and shear angles $\Delta \theta$ are scale-dependent \blue{on normalized CS thickness, $\lambda/\lambda_{p}$, where $\lambda_{p}$ is proton inertial length. The best power-law fits of the scale-dependence} of these quantities are given by Eqs. (\ref{eq:JJA_scale})--(\ref{eq:teta_scale}).
\end{enumerate}

The normalized properties of the proton kinetic-scale CSs at 0.2 au are very similar to those at 1 au. The CS properties are different in physical units due to different background plasma density and magnetic field, but in normalized units the CSs have similar and similarly scale-dependent current densities, amplitudes and shear angles. Based on observations and theoretical analysis at 1 au \citep{Vasko21a:apjl,Vasko21b:apjl}, we conclude that proton kinetic-scale CSs in the near-Sun solar are produced by turbulence cascade and are automatically in the parameter range, \blue{where reconnection cannot be suppressed by the diamagnetic suppression mechanism.}

\acknowledgments
The work of A.L. and Yu.K. was supported by the Swedish National Space Agency (SNSA) grant 206/19. The work of I.V. was supported by the Russian Science Foundation, grant No. 21-12-00416. The work of T.P. was supported by NASA Living With a Star grant \#80NSSC20K1781. The work of A.A. was supported by NASA Living With a Star grant \#80NSSC20K1788. \blue{The work of S.B. was supported by NASA Heliophysics Guest Investigator grant No. 80NSSC18K0646.} The data used in the analysis are publicly available at https://fields.ssl.berkeley.edu/data/.

\bibliography{full}

\end{document}